\documentclass[manuscript]{acmart}
\usepackage{tabularx}
\usepackage{booktabs}
\usepackage{multicol}
\usepackage{multirow}

\usepackage{amsmath,amsfonts,bm}

\def\eqref#1{equation~\ref{#1}}

\def\1{\bm{1}}

\DeclareMathAlphabet{\mathsfit}{\encodingdefault}{\sfdefault}{m}{sl}
\SetMathAlphabet{\mathsfit}{bold}{\encodingdefault}{\sfdefault}{bx}{n}

\newcommand{\ie}{\textit{i.e.,~}}
\newcommand{\eg}{\textit{e.g.,~}}

\usepackage[inline]{enumitem}
\setlist[description]{leftmargin=\parindent,labelindent=\parindent, font=\normalfont\itshape}

\usepackage{lipsum}
\usepackage{multirow}
\usepackage{tabularx}
\usepackage{bbm}
\usepackage{algorithm}
\usepackage[noend]{algpseudocode}
\usepackage{booktabs}
\usepackage{amsmath}

\usepackage{hyperref}
\usepackage{cleveref} 

\crefname{section}{\S}{\S} 
\crefname{subsection}{\S}{\S} 

\usepackage[rightcaption]{sidecap}

\AtBeginDocument{%
  \providecommand\BibTeX{{%
    \normalfont B\kern-0.5em{\scshape i\kern-0.25em b}\kern-0.8em\TeX}}}

\setcopyright{acmcopyright}
\copyrightyear{2018}
\acmYear{2018}
\acmDOI{10.1145/1122445.1122456}

\setcopyright{none}
\renewcommand\footnotetextcopyrightpermission[1]{}

\acmConference[Woodstock '18]{Woodstock '18: ACM Symposium on Neural
  Gaze Detection}{June 03--05, 2018}{Woodstock, NY}
\acmBooktitle{Woodstock '18: ACM Symposium on Neural Gaze Detection,
  June 03--05, 2018, Woodstock, NY}
\acmPrice{15.00}
\acmISBN{978-1-4503-9999-9/18/06}

\begin{document}

\title[Pre-trained Neural Recommenders]{Pre-trained Neural Recommenders: A Transferable Zero-Shot Framework for Recommendation Systems}
\newcommand{\name}{{\textsc{uni_rec}}}

\newtheorem{problem}{Problem}
\renewcommand{\algorithmicrequire}{\textbf{Input:}}
\renewcommand{\algorithmicensure}{\textbf{Output:}}
\newcommand{\subscript}[2]{$#1 _ #2$}

\author{Junting Wang}
\email{junting3@illinois.edu}
\affiliation{%
  \institution{University of Illinois, Urbana-Champaign}
  \streetaddress{P.O. Box 1212}
  \city{Dublin}
  \state{Ohio}
  \postcode{43017-6221}
}

\author{Adit Krishnan}
\email{junting3@illinois.edu}
\affiliation{%
  \institution{University of Illinois, Urbana-Champaign}
  \streetaddress{P.O. Box 1212}
  \city{Dublin}
  \state{Ohio}
  \postcode{43017-6221}
}

\author{Hari Sundaram}
\email{junting3@illinois.edu}
\affiliation{%
  \institution{University of Illinois, Urbana-Champaign}
  \streetaddress{P.O. Box 1212}
  \city{Dublin}
  \state{Ohio}
  \postcode{43017-6221}
}

\author{Yunzhe Li}
\email{junting3@illinois.edu}
\affiliation{%
  \institution{University of Illinois, Urbana-Champaign}
  \streetaddress{P.O. Box 1212}
  \city{Dublin}
  \state{Ohio}
  \postcode{43017-6221}
}

\setlength{\abovedisplayskip}{0.1cm}
\setlength{\belowdisplayskip}{0.1cm}
\setlength{\floatsep}{0.1cm}
\setlength{\textfloatsep}{0.1cm}
\setlength{\abovecaptionskip}{0.1cm}
\setlength{\belowcaptionskip}{0.1cm}
\setlength{\dbltextfloatsep}{0.1cm}
\setlength{\intextsep}{0.1cm}

\renewcommand{\shortauthors}{Trovato and Tobin, et al.}

\makeatletter
\newcommand{\algmargin}{\the\ALG@thistlm}
\makeatother

\begin{abstract}
Modern neural collaborative filtering techniques are critical to the success of e-commerce, social media, and content-sharing platforms. However, despite technical advances---for every new application domain, we need to train an NCF model from scratch. In contrast, pre-trained vision and language models are routinely applied to diverse applications directly (zero-shot) or with limited fine-tuning. Inspired by the impact of pre-trained models, we explore the possibility of pre-trained recommender models that support building recommender systems in new domains, with minimal or no retraining, without the use of any auxiliary user or item information. Zero-shot recommendation without auxiliary information is challenging because we cannot form associations between users and items across datasets when there are no overlapping users or items. Our fundamental insight is that the statistical characteristics of the user-item interaction matrix are universally available across different domains and datasets. Thus, we use the statistical characteristics of the user-item interaction matrix to identify dataset-independent representations for users and items.  We show how to learn universal (\ie supporting zero-shot adaptation without user or item auxiliary information) representations for nodes and edges from the bipartite user-item interaction graph. We learn representations by exploiting the statistical properties of the interaction data, including user and item marginals, and the size and density distributions of their clusters. 

With extensive experiments on five diverse public real-world datasets, we show that the proposed dataset-agnostic features, combined with a pre-trained recommendation model, generalizes to unseen users and unseen items within a dataset and across different datasets (\ie cross-domain, zero-shot) with comparable performance to state-of-the-art neural recommenders in traditional single-dataset settings. Furthermore, we show for the in-domain setting, the proposed features can boost the performance of existing state-of-the-art neural recommender models by up to 14\% on three out of five datasets via a simple post-hoc interpolation on the ranking prediction. Pre-trained models will have a significant application impact on developing recommender systems for new application domains with minimal fine-tuning (few-shot) or no training (zero-shot). Our code and datasets are available for review\footnote{\url{https://anonymous.4open.science/r/uni_recsys-2BFE}}.

\end{abstract}



\maketitle

\section{Introduction}
\label{sec:introduction}

Modern neural collaborative filtering (NCF) techniques are critical to the success of e-commerce, social media, and content-sharing platforms. However, the process of training these neural models has remained unchanged despite technical advances---for every new application domain, we need to train an NCF model from scratch. Further, even if we train the model for a domain, a large influx of new users and new items necessitates retraining for high performance. In contrast, in peer fields of AI, pre-trained vision models \cite{dosovitskiy2020image, he2016deep, radford2021learning}, and pre-trained language models \cite{bert, gpt3, raffel2020exploring, liu2019roberta} have transformed how practitioners use these models. In some scenarios \cite{gpt3, bert, he2016deep, raffel2020exploring}, a few fine-tuning steps are sufficient to achieve state-of-the-art performance (few-shot learning), while in other cases \cite{openai2023gpt4, gpt3}, practitioners use the pre-trained models directly without any fine-tuning (zero-shot learning). Thus in this paper, we ask: \textit{Can we develop pre-trained recommender models (PRMs) that support building recommender systems in new domains, with minimal or no retraining?} We want these models to generalize to 
\begin{enumerate*}
    \item unseen users and unseen items within the same dataset;
    \item new datasets that have no overlapping users, items, or auxiliary features with the training dataset.
\end{enumerate*}

At first sight, developing a pre-trained recommender model appears to be an impossible task, and conceptually distinct from developing a pre-trained language model. Language models learn transferable representations of a word conditioned on the past words within a text window to maximize the likelihood of the next word. Since the documents in the training corpus and the inference time applications share the same language(s), the models learn universal representations of a word and a conditional probability of occurrence given past words, which are then used to generate text or make inferences \cite{bert, gpt3}. In contrast, recommender system datasets only have two entities---users and items. Furthermore,  unlike documents from the same language(s) sharing the same set of words, where direct correspondences are possible, \textit{we cannot form correspondences between users and items across different recommendation datasets, when the sets of users, items or both may be disjoint}. In this work, we show how to develop fully transferable user and item representations that accomplish the goals of the pre-trained recommender model solely based on the user-item interaction matrix. Next, we briefly discuss the related approaches.


Our target application goals are related to research on zero-shot, cold-start, and cross-domain recommendations. Zero-shot recommendation approaches typically rely on shared auxiliary user or item information \cite{ding2021zero, feng2021zero, sileo2022zero}, \eg attributes and profiles. This information is typically unavailable or highly application-dependent, even when available. On the other hand, cross-domain recommender models are typically trained in information-rich source domains to improve the performance in information-sparse target domains \cite{man2017cross, hu2018conet, li2020ddtcdr}. This line of work typically assumes the existence of overlapping users or items across domains \cite{cao2022cross, li2020ddtcdr, emcdr}. Another approach is to exploit generic descriptions of users and items, \eg textual and visual descriptions, when tackling cross-domain recommendation with no user/item overlap \cite{liu2022collaborative, kanagawa2019cross} and cold-start users or items \cite{melu, dong2020mamo}. In this work, we develop a pre-trained recommender model for the zero-shot setting without relying on any shared auxiliary information.

\begin{figure*}
    \centering
    \includegraphics[width= \linewidth]{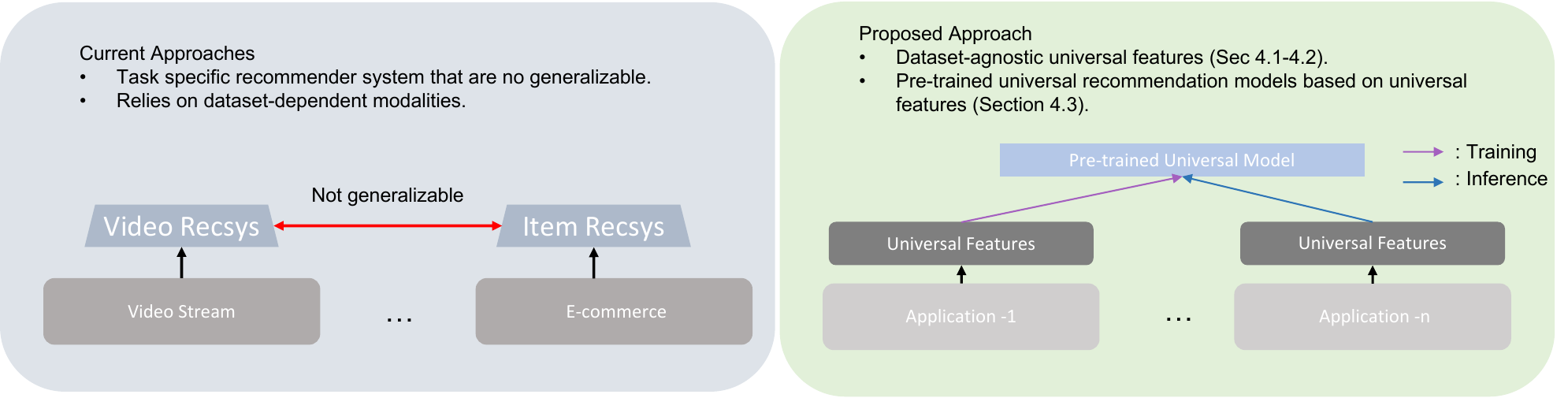}
    \caption{Current approaches v.s proposed approach. The current recommender systems use application-dependent information and are not generalizable to unseen datasets or new applications. Our proposed pre-trained recommender model effectively achieves cross-dataset generalization without using application-dependent auxiliary information. We first construct the dataset-agnostic universal features. Then, we pre-train the universal recommendation model on one dataset/application and perform zero-shot transfer to other domains.}
    \label{fig:intro}
\end{figure*}


\textbf{Present Work: } Our fundamental insight is that the statistical characteristics of the user-item interaction matrix are universally available across different domains and datasets. Thus, we hypothesize that the statistical characteristics of the user-item interaction matrix provide a strong clue toward dataset-independent representations for users and items. We know from seminal work by~\citet{Barabasi1999} that the marginal distributions of the user interactions on social networks form a heavy-tailed distribution and that these distributions arise due to the finite cognitive constraints of the users~\cite{Barabasi2005}. While the notion of popularity bias is well-known to the recommender systems community~\cite{Abdollahpouri2020}, early experiments by~\citet{Salganik2006} establish that markets \textit{will develop} a similar heavy-tailed distribution for items over time, under a canonical model of item recommendations. Intuitively, across datasets, predicting how a highly (or less) active user interacts say with popular (or sparse) items should be predictable from the statistics, and to a large degree, independent of the descriptions of the user and the item.

Our goal is to learn representations of user $u$ and item $v$ from the marginals (\ie user and item activity) and the representation of their interactions $(i,j)$ from the joint distribution of the user-item interaction matrix. We represent the user and item interaction matrix as a bi-partite graph and proceed as follows (details in~\cref{sec:methods}): 
\begin{itemize*}[font=\normalsize\itshape]
    \item[Activity features: ] From the user and item interaction volume, we identify the bin (\eg quartile) that each user ($u$) and item ($v$) falls into and construct one-hot activity bin representations. Then, we compute activity features of each user or item as the sum of the one-hot activity bin representations of its neighbors.
    \item[Co-occurrence features: ] We apply Node2Vec\cite{grover2016node2vec} to the bipartite interaction graph and cluster the node embeddings. We then bin these clusters based on their sizes and densities and compute one-hot vectors for each node (user/item node) based on the size and density bins of the clusters they belong to. Finally, we compute the final node representations as weighted aggregates of the one-hot vectors of their neighboring nodes.
    \item[Interaction features: ] We use the user and item Node2Vec embeddings to compute embeddings for each edge $(i,j)$ in the interaction graph. Similar to the co-occurrence representations, we cluster these edge embeddings and compute the size and density distributions of these clusters. We use these distributions to compute one-hot vectors for each edge. The final interaction representation of each node is the aggregate of the one-hot vectors of its edges.
\end{itemize*}


To summarize, our key contributions are as follows: 
\begin{description}
    \item[Pre-trained recommender model: ] Unlike standard neural recommenders \cite{bpr, ncf, ngcf, lightgcn} which must be trained on each new dataset, we are the first to explore the possibility of pre-trained recommender systems. We accomplish this by learning representations of users and items from the statistical properties of the interaction matrix. A pre-trained model will have a significant application impact as we can use it to develop a recommender system for a new application domain with minimal fine-tuning (few-shot) or no training (zero-shot).
    
    \item[Universal user, item, and interaction representations: ] We show how to learn universal (\ie supporting zero-shot adaptation without user or item auxiliary information) representations for nodes and edges from a bipartite user-item interaction graph. In contrast, prior works on zero-shot recommendation or cold-start~\cite{melu, dong2020mamo, ding2021zero, feng2021zero} assume the presence of additional auxiliary information while cross-domain transfer~\cite{emcdr, catn, zhu2021transfer} assumes overlapping users or items. We learn representations by exploiting the statistical properties of the interaction data, including user and item marginals, size and density distributions of clusters. These representations are crucial to zero-shot knowledge transfer without any auxiliary information or overlapping users or items.
    
\end{description}

 Through extensive experiments on five real-world datasets, we demonstrate the generalizability of the proposed set of universal features to unseen users and items, as well as new datasets. In addition, we empirically show that they can improve the performance of state-of-the-art neural collaborative filtering methods via post-hoc interpolation by up to 14 \% on three of the five selected datasets.

\section{Related Work}
\label{sec:related_work}

In this section, we briefly review a few related lines of work that is relevant to this paper.

\textbf{Neural Collaborative Filtering}: The core idea of latent-factor collaborative filtering (CF) methods is to learn latent representations of users and items to maximize the likelihood of observed user-item interactions. Neural CF (NCF) methods enhance the capability of traditional CF methods with non-linear latent factor interactions - e.g., via interaction modeling \cite{bpr, ncf, lrml}, or graph convolution on the interaction graph~\cite{ngcf, lightgcn}. However, their recommendation results rely heavily on historical interactions between users and items. While some NCF methods handle unseen items or unseen users \cite{vae-cf, protocf} when they interact with the existing users or items, they cannot leverage the interaction between unseen users and unseen items without model retraining. Therefore, neural collaborative filtering methods cannot generalize to new data from the same domain or unseen data across domains.

\textbf{Cross-domain Recommendation}: The general goal of cross-domain recommendation is to leverage relatively data-rich domains to improve the recommendation performance of the data-sparse domain\cite{hu2018conet, li2020ddtcdr, man2017cross}. However, most of this work assumes that there are overlapping users and items \cite{emcdr, catn, zhu2021transfer} for effective knowledge transfer and domain adaptation.
   \begin{figure}[t]
     \centering
      \includegraphics[width=0.5\linewidth]{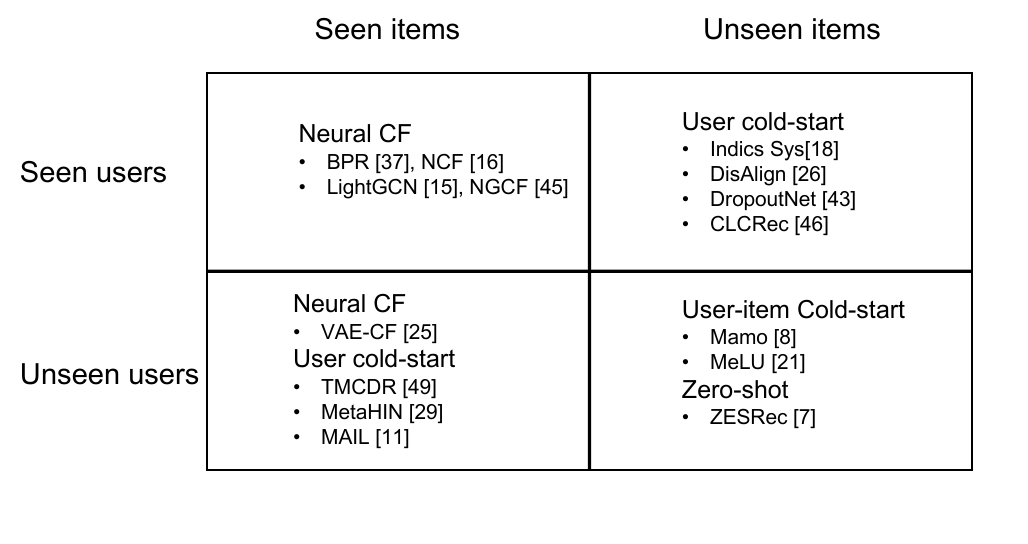}
      \caption{Categorization of related literature.}
      \label{fig:activity_distribution}
  \end{figure}
  
\textbf{Cold-start Recommendation}: 
Methods have been proposed to tackle either the cold-start item case \cite{dropoutnet, wei2021contrastive, du2020learn, liu2021leveraging, huan2022industrial} or the cold-start user case \cite{lu2020meta, zhu2021transfer, feng2021zero}. However, they rely on auxiliary information to generate the initial representation and can only handle one-sided cold-start problems. Some past work explores meta-learning for cross-domain recommendation tasks on unseen users and unseen items \cite{melu, dong2020mamo}. However, they rely on side information to learn the representations, which is application dependent. Note that the primary goal of this paper is not to address the cold-start recommendation problem. However, we pursue a highly generalizable recommender system that can work on any new domain with just the interaction matrix.

\textbf{Zero-shot Learning}: Zero-shot learning originated in computer vision to make predictions on previously unseen test classes~\cite{wang2019survey, feng2021zero}. A general approach is to make image predictions based on the images' semantic information. Analogously, some zero-shot approaches on recommender systems try to map the attribute space of users and items into their latent space \cite{ding2021zero, feng2021zero, melu, dong2020mamo}. However, they all rely on highly dataset-dependent features. In this work, we only look at the universal data present in all recommendation datasets: the user-item interaction matrix. We identify dataset-agnostic features purely based on user-item interactions, and we show that these features can effectively help us generalize recommender systems to unseen users and items within the same domain or across different datasets. 

Note that, unlike prior related work that relies on overlapping entities or side information, we view the statistical properties of the interaction matrix as a universal information source independent of the dataset. Thus, we aim to transfer models built on these universal features. To the best of our knowledge, we are the first to explore pre-trained recommender models and tackle the problem of zero-shot recommendations to unseen users and items without any auxiliary information or common entities across domains.

\section{Problem Definition}
\label{sec:problem_formulation}

\begin{table}[t]
    \centering
    \begin{tabular}{rl}
        \toprule
        Symbol                & Description \\
        \midrule        
$k$ & number of bins/feature dimension  \\
$\mu_U$, $\mu_V$ & sorted activity distribution of users and items\\
$Q_u$, $Q_v$& users' and items
 one-hot encoded activity bin representation\\
$\mathcal{N}_u$, $\mathcal{N}_v$ & neighboring nodes of user $u$ and item $v$ in the user-item bi-partite graph \\
$A_u$ , $A_v$ & activity-based feature of users and items respectively\\
$C_{U, D}$, $C_{V, D}$ & density-based co-occurrence feature of users and items\\
$C_{U, S}$, $C_{V, S}$ & size-based co-occurrence feature of users and items\\
$E_{U, D}$, $E_{V, D}$ & density-based interaction feature of users and items\\
$E_{U, S}$, $E_{V, S}$ & size-based interaction feature of users and items\\
$g_U$, $g_V$ &Node2Vec embeddings of users and items \\

$t_U$ &user clusters centroids\\
$t_I$ & item clusters centroids\\
$d(\cdot, \cdot)$ & bounded distance metric between two vectors\\
$s(\cdot, \cdot)$ & similarity metric between two vectors\\

        \bottomrule
    \end{tabular}
    \caption{Notation}
    \label{tab:notations}
\end{table}
In this paper, we consider our recommendation problems with only implicit feedback, \ie no explicit ratings or auxiliary information. We denote the user set as $\mathcal{U}$, the item set as $\mathcal{V}$, and the interaction matrix as $\mathcal{M} = |\mathcal{U}| \times |\mathcal{V}|$. Note that the interaction matrix also allows for a bipartite graph formulation, with nodes being users or items and where an edge implies an observed interaction. Neural collaborative filtering methods essentially a train recommender system $\mathcal{R}$ that learns the latent representations for every user $u \in \mathcal{U}$ and item $v\in \mathcal{V}$. The overall goal of $\mathcal{R}$ is to learn a personalized scoring function $f(v | u, \mathcal{M})$ that captures the likelihood of $v $ interacting with $u$. 

In this paper, we consider the following three recommendation problem formulation: 

\begin{problem}[{\textbf{In-domain Recommendation}}] 
Given an interaction matrix $M$ over $\mathcal{U}$ and $\mathcal{V}$, learn a recommender system $\mathcal{R}$ that produce a scoring function $f(v |u, \mathcal{M})$ for $v \in \mathcal{V}$ and $u \in \mathcal{U}$.
\end{problem}
\begin{problem}[{\textbf{Zero-shot In-domain Recommendation}}] 
Given an interaction matrix $M'$ over $\mathcal{U}'$ and $\mathcal{V}'$, where $M'$ and $M$ belong to the same domain, $\mathcal{U}' \cap \mathcal{U} = \varnothing $, and $\mathcal{V}' \cap \mathcal{V} = \varnothing $, produce a scoring function $f'(v'|u', \mathcal{M}')$ for $v' \in \mathcal{V}'$ and $u' \in \mathcal{U}'$ without training the scoring function on $M'$.
\end{problem}
\begin{problem}[{\textbf{Zero-shot Cross-domain Recommendation}}] 
Given an interaction matrix $M^*$ over $\mathcal{U}^*$ and $\mathcal{V}^*$, where $M^*$ and $M$ belong to different domains, $\mathcal{U}^* \cap \mathcal{U} = \varnothing $, and $\mathcal{V}^* \cap \mathcal{V} = \varnothing $, produce a scoring function $f^*(v^*|u^*, \mathcal{M}^*)$ for $v^* \in \mathcal{V}^*$ and $u^* \in \mathcal{U}^*$ without training the scoring function on $M^*$ directly.
\end{problem}

Note that, unlike prior works in cross-domain and cold-start recommendation, our goal is to investigate universal properties of users and items that are shared within the dataset and across the dataset without the help of auxiliary information.

\section{Pre-trained Recommender System Framework}
\label{sec:methods}
Prior work on zero-shot recommendations leverage application-dependent features (\eg auxiliary use features, item descriptions) to make recommendations. In contrast, we leverage the universal statistical characteristics of the user-item interaction matrix and develop a generalizable pre-trained recommender system with these features. In this section, we first introduce three application and data-independent distributional features purely based on the interaction matrix. Specifically, the user/item activity distribution (\cref{sec:act_f}), node representations from the user/item co-occurrence graph (\cref{sub:Nodal representations co-occurence}), and edge representations from user-item interactions (\cref{sec:int_f}). Then, we use these features to develop our transferable pre-trained recommendation model (\cref{sec:model_details}).
   \begin{figure*}[t]
     \centering
      \includegraphics[width=\linewidth]{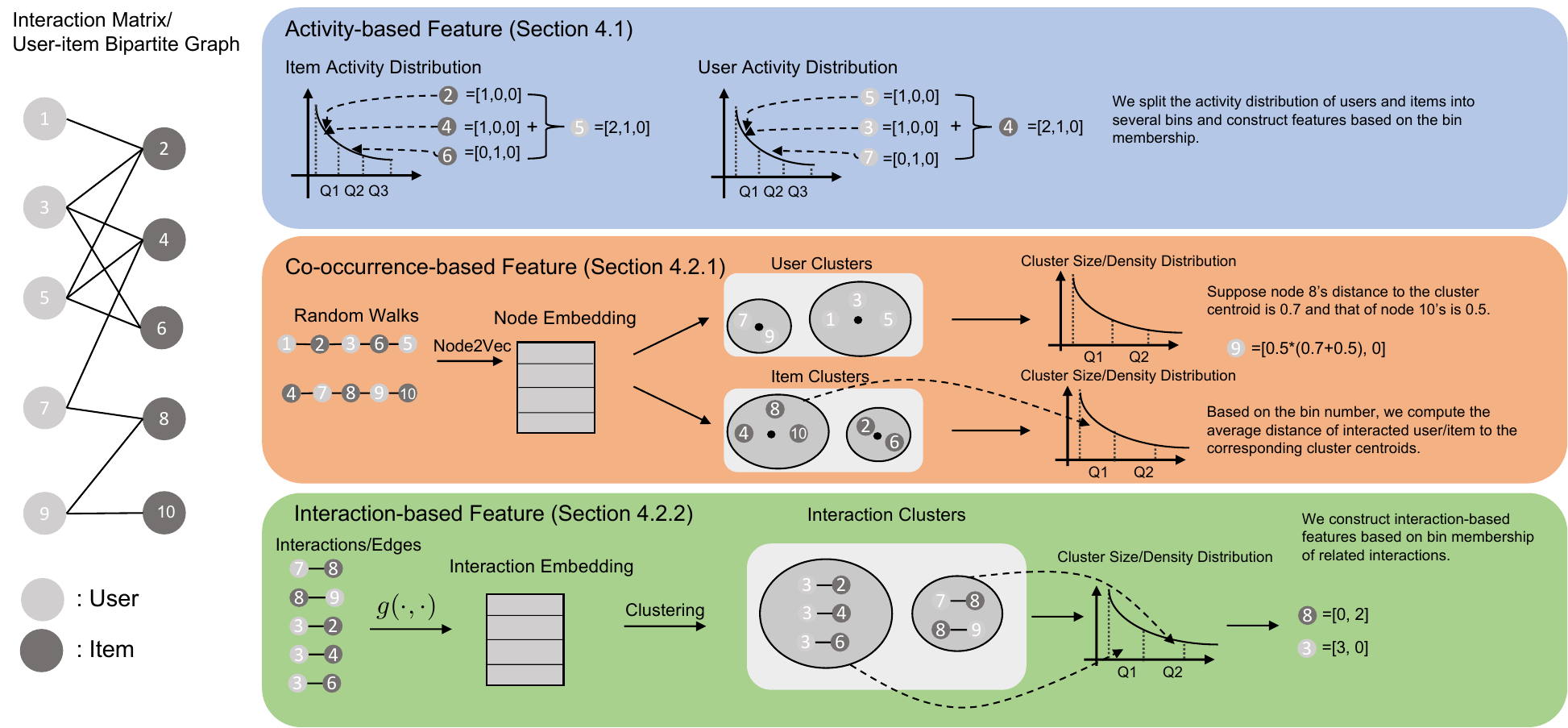}
      \caption{Feature construction process. We provide an example of constructing features based on a simplistic user-item interaction matrix.}
      \label{fig:model}
  \end{figure*}
\subsection{Marginal Activity-based Features}
\label{sec:act_f}
E-commerce platforms rely on users' behavior history to make recommendations. We know from work by~\citet{Barabasi1999} that the user activity distribution follows a heavy-tailed distribution, \ie the number of users who interact with a certain number of items follows a power law distribution. These distributions arise out of human cognitive constraints~\cite{Barabasi2005} and are not application-dependent. Furthermore, experimental work by~\cite{Salganik2006} reveals that the item activity distribution also becomes progressively unequal (they study music) under a canonical model of item recommendation. Thus prior work strongly suggests that the activity distribution of users and items in \textit{across} recommendation datasets follow similar patterns. 

Identifying where the user or item lies in the activity distribution is predictive of their future interactions, independent of the application. For example, suppose Alice and Bob are two users across two application domains (\eg Netflix and Amazon). Then, we conjecture that their interaction histories in terms of activity distribution help us determine if they are similar. Specifically, do they interact with mostly popular items or long-tail items? If Alice and Bob's interaction histories primarily contain popular items in their respective domains, then they are more similar to each other than to other than users who interact more with long-tail items (items with fewer interactions). Analogously, items that are purchased mostly by highly active users are different from items that are purchased by long-tail users. The transferred knowledge on cross-domain similarity is \textit{not} based on the specific descriptions of the users or items, but rather on the statistical characteristics of their interaction histories, unlike past work~\cite{ding2021zero, feng2021zero, dong2020mamo, melu}. We first categorize users and items by their position in the activity distribution and then construct features that describe their interaction histories, \ie the activity level of their interacted items and users respectively.



Formally, given an interaction matrix $\mathcal{M}$, to construct activity-based user features, we first calculate the sorted (either from largest to smallest or reverse) activity distribution $\mu_V$ of the item set $\mathcal{V}$ based on the number of users each item has interacted with. Then, we split $\mu_V$ into $k$ different bins. Next, we represent each item $v_j$ by its one-hot encoded bin representation vector $Q_{v_j}$. For example, given $k=3$ bins, item $v_j$ in activity bin two is embedded as $Q_{v_j} = [0, 1, 0]$ (see~\Cref{fig:model}). We perform similar procedure for the user side as well. The activity-based features of user $u_i$ and $v_j$ are constructed as:  
\begin{equation}
A_{u_i} = \sum_{v_j \in \mathcal{N}_{u_i}} {Q_{v_j}}, \quad A_{v_j} = \sum_{u_i \in  \mathcal{N}_{v_j}} {Q_{u_i}}
 \label{eqn:ad_u}
\end{equation}

Note that in~\Cref{eqn:ad_u}, the aggregate representation of each user counts the number of interacted items in each activity bin and vice-versa (\ie the neighboring nodes). We find that the un-normalized raw sum provides the best performance. Next, we discuss joint features extracted from the interaction histories of users and items.

\subsection{Joint Interaction-based features}
\label{sec:co_f}
Apart from learning the latent factors of users and items from their marginal activity distributions, an alternative way to measure similarities of users and items is through co-occurrence/joint distribution~\cite{cooccur}. This intuition follows literature in learning contextual word embedding~\cite{word2vec}. Items that frequently co-occur in different users' interaction histories should be similar, and vice-versa. Since similar items have closer embeddings, their representations form clusters in the latent space, which may reflect a contextual factor of items such as item category or price. To capture the joint distribution of users and items, we apply Node2Vec\cite{grover2016node2vec} to the bipartite user-item interaction graph to get the user embeddings $g_U$ and item embeddings $g_V$.


\subsubsection{Co-occurrence based node representations}
\label{sub:Nodal representations co-occurence}

 We cluster these embeddings $g_U$ and $g_I$, into $m$ and $n$ clusters, respectively. We denote the clusters of users and items as $L_U = \{L_U^1 \dots L_U^m\}$ and $L_V = \{L_V^1 \dots L_V^n\}$ respectively. One critical challenge of constructing universal features based on co-occurrence clusters is that clusters from different datasets might have drastically different structures and sizes. Therefore, we propose constructing features based on clusters' fundamental characteristics, \ie size, and density. 
We characterize user clusters and item clusters by computing the size $S$ and the density $D$ of each cluster as follows:
\begin{equation}
    S_U^p = |L_U^p|,\quad S_V^q = |L_V^q|, \quad D_U^p = \frac{1}{|L_U^p|}\sum_{u \in L_U^p} d(u, t_U^p) ,\quad D_V^q = \frac{1}{|L_V^q|}\sum_{v \in L_V^q} d(v, t_V^q),
    \label{eqn:density}
\end{equation}
where, $d(\cdot, \cdot)$ can be any bounded distance metric between two vectors, \eg cosine distance and where, $t_V^q$ and $t_U^p$, represent the cluster centroids of the $q$-th item and $p$-th user cluster respectively.

Then, similar to the activity distribution-based feature, we split the cluster size $S$ and density $D$ distributions into $k$ bins. Note that size and density distribution can be split into different numbers of bins. We denote $R_{U, S}^{i}$ and $R_{U, D}^{i}$ as the one-hot encoded bin membership of user $u_i$ based on the size and density respectively. Similarly we can compute one-hot encoded bin membership $R_{V, S}^{j}$ and $R_{V, D}^{j}$ for item $v_j$.

For the \textit{user} co-occurrence feature, we compute how close are the items that the user interacts with to their corresponding cluster centroids. The user representation is then the weighted sum of the bin membership vectors of the items that they interact with. Thus, we have a user representation vector derived from joint interaction, one for size and one for density. Formally, for a given user $u_i$: 

Let $Z_{U, D}^{i} = (\mathrm{diag}[\sum_{v_j \in \mathcal{N}_{u_i}} R_{V,D}^j])^{-1}$ and $Z_{U, S}^i = (\mathrm{diag}[\sum_{v_j \in \mathcal{N}_{u_i}} R_{V,S}^j])^{-1}$, where $\mathrm{diag(\cdot)}$ creates a diagonal matrix with the given vector.

\begin{equation}\label{eqn:cu_feature}
\begin{split}
C_{U, D}^i = {Z_{U, D}^i}\sum_{v_j \in \mathcal{N}_{u_i}} R_{V,D}^j\cdot s(g_V^j, t_V^j),\quad
C_{U, S}^i = {Z_{U, S}^i}\sum_{v_j \in \mathcal{N}_{u_i}} R_{V,S}^j\cdot s(g_V^j, t_V^j)
\end{split}
\end{equation}
where $s(\cdot, \cdot)$ is any similarity metric measuring similarities of vectors, \eg cosine similarity, and where $t_V^j$ is the corresponding cluster centroid for item $v_j$. For items, we follow the same procedure to compute the co-occurrence-based features. For the \textit{item} co-occurrence feature, we compute how close are the users that the item is seen interacting with to their corresponding cluster centroids. The item representation is then the weighted sum of the bin membership vectors of the users that they interact with. 

\begin{equation}\label{eqn:ci_feature}
\begin{split}
C_{V,D}^j = Z_{V, D}^{j}\sum_{i \in \mathcal{N}_{v_j}} R_{U, D}^i\cdot s(g_U^i, t_U^i) ,\quad
C_{V,S}^j = Z_{V, S}^{j}\sum_{i \in \mathcal{N}_{v_j}} R_{U, S}^i\cdot s(g_U^i, t_U^i)
\end{split}
\end{equation}

To summarize, we cluster the user and item embeddings derived from the interaction matrix to compute size and density features for users and items (\ie the nodal features), which are derived from the cluster size and density distributions. As a reminder, we focus on cluster size and density distribution membership since user-item interaction clusters across different datasets can vary widely. Next, we discuss how we use the co-occurrence user embeddings $g_U$ and item embeddings $g_I$ to learn edge embeddings for each edge in the graph.

\subsubsection{Edge representations, Interaction-based features}
\label{sec:int_f}


We compute the edge representations $E$, in a manner similar to prior work~\cite{grover2016node2vec}, via a binary operator $b(\cdot, \cdot)$ on the learned  node embeddings $(g_U^i, g_V^j)$ where the binary operator is, for example, a weighted Hadamard product: 
\begin{equation}
E_{i,j} = b(g_U^i, g_V^j) 
 \label{eqn:e_i,j}
\end{equation}
Similar to the node features, we cluster the edge representations into $m$ clusters, where each of the clusters may represent a particular type of edge/interaction. Like the node clusters, the edge clusters might also vary across different datasets. Therefore, we characterize these edge clusters as $L_E = \{L_E^1 \dots L_E^m\}$,  based on their cluster size $S_E$ and density $D_E$. Then, we split the distribution of cluster size and cluster density into $k$ bins. Similar to the activity-based features, we represent each interaction $m_{i,j} \in \mathcal{M}$ between user $u_i$ and item $v_j$ by its one-hot encoded size bin representation vector ${P}_S^{i,j}$ and density bin representation vector ${P}_D^{i,j}$. Then, we compute the interaction-based features of user $u_i$ as follows: 

\begin{equation}
{E_{U, S}}^i = \sum_{j \in \mathcal{N}_{u_i}} {P}_S^{i,j}, \quad {E_{U, D}}^i = \sum_{j \in \mathcal{N}_{u_i}} {P}_D^{i,j}
 \label{eqn:eus}
\end{equation}

Similarity, we compute the interaction-based feature of item $i_j$ as: 
\begin{equation}
{E_{V,S}}^j = \sum_{i \in \mathcal{N}_{v_j}} {P}_S^{i,j}, \quad {E_{V,D}}^j = \sum_{i \in \mathcal{N}_{v_j}} {P}_D^{i,j}
 \label{eqn:eis}
\end{equation}

Again notice in~\Cref{eqn:eus,eqn:eis} that the edge representations corresponding to users are the sum of the bin membership vectors of the items that they interact with, and vice-versa.


   \begin{figure}[t]
     \centering
      \includegraphics[width= 80mm]{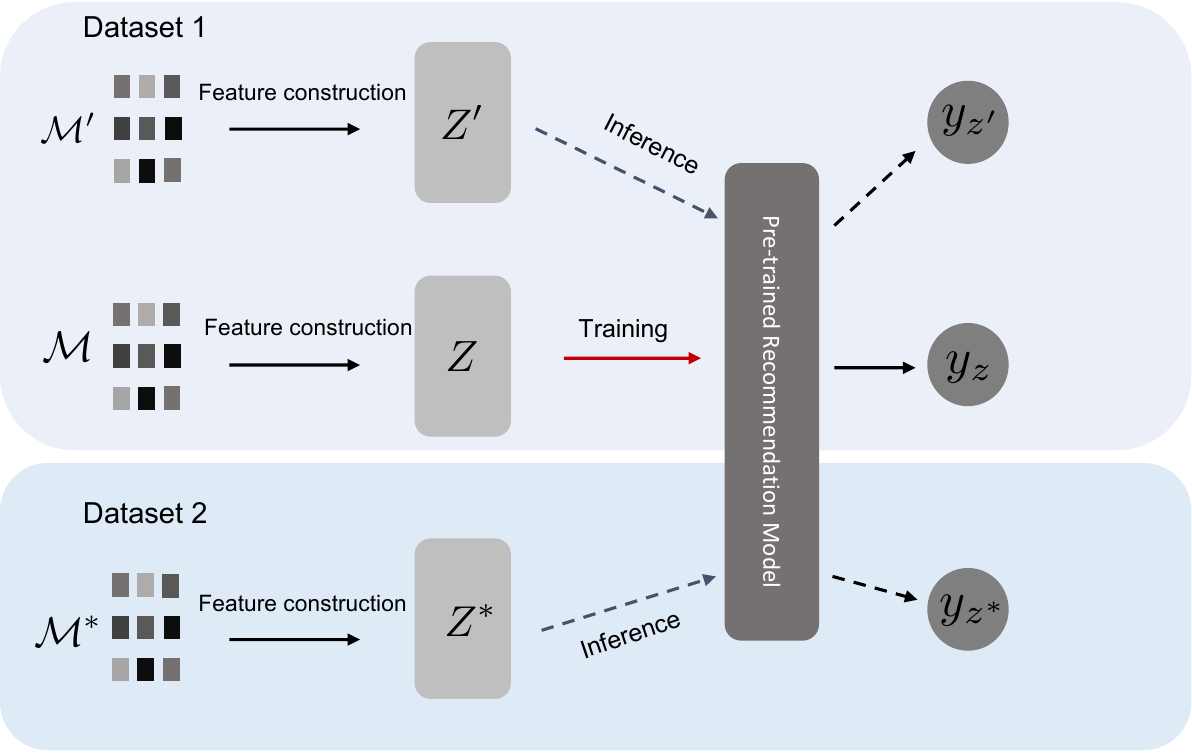}
      \caption{Training and inference process. Not that we only use the interaction matrix $\mathcal{M}$ for training. Both $\mathcal{M'}$ and $\mathcal{M^*}$ are only used for the inference. We introduce the evaluation detail in section ~\ref{sec:setup}.}
      \label{fig:model}
  \end{figure}

\subsection{Pre-trained model details}

Now we show how we can use the above features derived from the interaction matrix to develop a pre-trained model that can be deployed in a zero-shot scenario. We begin with model training, then show how we can use the pre-trained model for inference.

\label{sec:model_details}
\subsubsection{Training with dataset-independent features}
In the preceding sections, we have shown how we can derive features for users and items from the interaction matrix. Assume that the user features are denoted as $f_u$ and the item features are denoted as $f_v$; for example, these could be the activity-based features discussed in~\cref{sec:act_f} or the interaction-based features discussed in~\cref{sec:int_f}.

Given features $f_u, f_v$, we train two separate Multi-layer Perceptrons (MLP) as the pre-trained model for a given feature $f$ on users' and items', denoted as $MLP^U_f$ and $MLP^V_f$, respectively. If we represent the output of  $MLP^U_f(f_u)$ as a vector $e_{f,u}$ and the output of $MLP^V_f(f_v)$ as a vector $e_{f,v}$, then we can compute the prediction of the interaction $\hat{y}_f(u, v)$ between user $u$ and item $v$ given feature pair $(f_u, f_v)$ as their inner product:

\begin{equation}
\hat{y}_f(u, v) = <e_{f,u}, e_{f,v}>
 \label{eqn:pred}
\end{equation}

We employ the \textit{Bayesian Personalized Ranking}(BPR) loss proposed in \cite{bpr} as our objective function to train the MLPs, with the loss $L$ defined as: 
\begin{equation}
%
L = - \sum_{u \in \mathcal{U}}\sum_{v^{+}\in\mathcal{N}_u} \sum\limits_{ v^- \notin \mathcal{N}_u} \ln \sigma(\hat{y}_f(u, v^+) - \hat{y}_f(u, v^-) )
 \label{eqn:loss}
\end{equation}
We use Adam \cite{kingma2014adam} as the optimizer in a mini-batch manner.

\subsubsection{Model Inference}
Our inference setup varies depending on the recommendation problem introduced in section \ref{sec:problem_formulation}. For problems where we want to infer the unseen users and unseen items, \ie zero-shot in-domain recommendation and zero-shot cross-domain recommendation, we construct the universal features on the unseen interaction matrix and directly apply the model pre-trained on the seen interaction matrix to get the prediction. Formally, given a universal feature pair $f_{u'}$ for the unseen user and $f_{v'}$ for the unseen item, and pre-trained $MLP^U_f$ and $MLP^V_f$, we compute user and items representations as $e_{f,u'} = MLP^U_f(f_{u'}), e_{f,v'}=MLP^I_f(f_{v'})$ prediction as their inner product: 
\begin{equation}
\hat{y}_f(u', v') = <e_{f,u'}, e_{f,v'}>
 \label{eqn:unseen_pred}
\end{equation}
In addition, since the proposed dataset-agnostic features are complementary to each other, we propose a simple interpolation approach to compute the feature-based prediction: 
\begin{equation}
\hat{y}_z(u, v) = \alpha \hat{y}_a(u, v) + \beta \hat{y}_c(u, v) + \gamma \hat{y}_i(u, v)
 \label{eqn:feature_interpolate}
\end{equation}
where $\hat{y}_a(u, v)$, $\hat{y}_c(u, v)$, and $\hat{y}_i(u, v)$ represent the activity-based, co-occurrence-based, and interaction-based predictions respectively. We impose constraint on the range of $\alpha$, $\beta$, and $\gamma$, where $\alpha,\beta,\gamma \in (0,1)$ and $\alpha + \beta + \gamma = 1$. Since both the co-occurrence-based and interaction-based features can be constructed by either size or density. We aggregate the size feature and density feature output via an interpolation based on $\delta, \epsilon \in (0,1)$. Formally, 

\begin{equation} \label{eqn:feature_mean}
\begin{split}
    \hat{y}_c(u, v) &= \delta\hat{y}_c^s(u, v) + (1-\delta)\hat{y}_c^d(u, v)\\
\hat{y}_i(u, v) &= \epsilon\hat{y}_i^s(u, v) + (1-\epsilon)\hat{y}_i^d(u, v)
\end{split}
\end{equation}
where $\hat{y}_c^s(u, v)$ and $\hat{y}_c^d(u, v)$  are the co-occurrence-based features constructed based on size and density respectively, similar for $\hat{y}_i^s(u, v)$ and $\hat{y}_i^d(u, v)$. We develop these simple aggregations to help understand the role of each type of feature in predictions. We note that we can always train two MLPs, one for users and the other for items, by concatenating the statistical features discussed in this section to further improve the performance.

For the in-domain recommendation scenario, often, the application would already be working with a trained NCF model. In this case, we could still boost the performance by incorporating the proposed statistical features. To do so, we compute the features on the training set and learn the MLP layers via equation~\cref{eqn:loss}. Since the proposed feature set focuses on the zero-shot recommendation on unseen users and items, we propose to ensemble the predictions from $\hat{y}_z(u, v)$ and any NCF models. One simple interpolation approach is given by: 
\vspace{-5pt}
\begin{equation}
\hat{y}(u,v) =  (1 - \eta) \cdot \hat{y}_z (u,v) + \eta  \cdot \hat{y}_b (u,v)
 \label{eqn:interpolation}
\end{equation}
where $\eta \in (0,1) $ and $\hat{y}_b (u,v)$ represent the output of other NCF models.

To summarize, in this section, we showed how to develop a pre-trained recommender model based on statistical features of the interactions between users and items. We are motivated by strong empirical~\cite{Barabasi1999,Barabasi1999} and experimental work~\cite{Salganik2006} that user behaviors will manifest as statistical invariants across interaction datasets. We identified representations for users and items via marginal ~\cref{sec:act_f} and joint~\cref{sec:co_f} distributions of the interactions. Then, in \cref{sec:model_details}, we showed how to learn the model, and how to use the model to make predictions in different recommendation scenarios. We will evaluate the proposed model in the next section.

\section{Experiments}
\label{sec:experiments}

Here, we present extensive experiments on five real-world datasets to evaluate the effectiveness of the proposed universal features and the pre-trained recommendation framework under different settings. We introduce the following research questions (RQ) to guide our experiments.
\begin{enumerate*}[label={RQ\arabic*}:]
\item How well can proposed universal features perform on the in-domain recommendation task?
\item Can the universal features be used for the zero-shot in-domain recommendation problem?
\item Can the universal features to generalize across datasets?
\item What is the effect of the amount of training data available?
\item How will the feature parameters impact the performance? 
\end{enumerate*}

\vspace{-5pt}

\subsection{\textbf{Datasets} }
\label{sec:datasets}
We conducted experiments on five publicly available benchmark datasets from different domains and applications.
\begin{description}[font=\mdseries\scshape]
\item[Epinions]\footnote{https://www.cse.msu.edu/tangjili/datasetcode/truststudy.html}:  product ratings from Epinions; we binarize explicit ratings by keeping ratings of three or higher.
\item[Yelp]\footnote{https://www.yelp.com/dataset}: user ratings on local businesses located in the state of Arizona,  obtained from Yelp dataset challenge.
\item[MovieLens 1M]\footnote{\url{https://grouplens.org/datasets/movielens}}: movie ratings dataset from the MovieLens website.
\item[Douban Movie]: movie review dataset collected from Douban\footnote{\url{www.douban.com}} by \cite{song2019session}.
\item[Amazon-Sport]\footnote{\url{https://nijianmo.github.io/amazon/index.html}}: sport category of the amazon product review dataset collected by \cite{ni2019justifying}.
\end{description}

 \begin{table}[t]
 \centering
 \small
 \begin{tabular}{@{}p{0.23\linewidth}p{0.11\linewidth}p{0.11\linewidth}p{0.13\linewidth}p{0.11\linewidth}p{0.11\linewidth}@{}}
 \toprule
 \textsc{Dataset}  &\textsc{Epinions} & \textsc{Yelp}  & \textsc{MovieLens} &\textsc{Douban}  & \textsc{Sport}\\
 \midrule
 \# Seen Users &5,485 & 9,481 & 3,013&23,218&20,865\\
  \# Seen Items &5,469 & 5,644 & 1,557&12,452&10,609\\
  \# S-S Interactions & 73,459 & 133, 227 & 211,059&1,777,100&153,223\\
   \# Unseen Users & 5,699 & 9,585& 3,004&23,286&22,343\\
    \# Unseen Items&5,613 & 5,549& 1,571&12,500&11,292\\
    \# U-U Interactions & 77,838 & 136,557& 206,096& 1,801,347&165,836\\
 \bottomrule
 \end{tabular}
 \caption{Dataset statistics}
\label{tab:stats}
\end{table}
For the preprocessing step, first, we randomly partition the user and item set from each dataset into two equally sized parts, \ie seen users and unseen users, seen items and unseen items. Then, we select the corresponding seen users-seen items and unseen users-unseen items to construct the dataset for in-domain recommendation and zero-shot in-domain recommendation respectively. We preprocess each partitioned dataset under the 5-core setting~\cite{10-core}, to retain users and items with at least five interactions (Table~\ref{tab:stats}). In addition, for each user, we select 70\% of its interaction to construct the training set, and the remaining 30\% constitute the test set. From the training set, we randomly select 10\% of the interactions as the validation set for hyper-parameter tuning and construct the proposed features for each user and item based on the remaining interaction in the training set. We use the training interactions for the second and the third recommendation problems in ~\cref{sec:problem_formulation} just for feature construction. The reason why we split the data by percentage is to maximally preserve the original data distribution.




\subsection{\textbf{Baselines}}
\label{sec:baselines}
We adopt four state-of-the-art collaborative filtering methods to benchmark the in-domain recommendation task results. In addition, to the best of our knowledge, we are the first to tackle the zero-shot learning setting with no side information. Therefore, for the zero-shot recommendation problems, we adopt an item popularity-based method to produce the rank list by ranking the items based on their interaction counts. 
\begin{description}[font=\mdseries\scshape]
\item[BPR~\cite{bpr}:] neural CF model with matrix factorization and Bayesian Personalized Ranking loss.
\item[NCF ~\cite{ncf}:] neural CF model with non-linear neural layers between the user and item latent embeddings.
\item[NGCF ~\cite{ngcf}:] graph-based NCF model with embedding propagation layers on the user-item interaction graph.
\item[LightGCN ~\cite{lightgcn}:] state-of-the-art graph neural network models for recommendation with simplified GNN designs. 
\item[MostPop \cite{bpr, ncf, kanagawa2019cross}:] create rank lists by ranking items based on their interaction counts.
\end{description}
\newcommand*{\factor}{0.034}
\subsection{\textbf{Experimental Setup}}
\label{sec:setup}
\begin{table*}[t]
\centering
 \small
\begin{tabular}{p{0.12\linewidth}p{\factor\linewidth}p{\factor\linewidth}p{\factor\linewidth}p{\factor\linewidth}p{\factor\linewidth}
p{\factor\linewidth}p{\factor\linewidth}p{\factor\linewidth}p{\factor\linewidth}p{\factor\linewidth}p{\factor\linewidth}
p{\factor\linewidth}p{\factor\linewidth}p{\factor\linewidth}p{\factor\linewidth}}
\toprule
\textsc{Dataset} & \multicolumn{3}{l}{\textsc{Epinions}} & \multicolumn{3}{l}{\textsc{Yelp}} & \multicolumn{3}{l}{\textsc{MovieLens}}& \multicolumn{3}{l}{\textsc{Douban}}& \multicolumn{3}{l}{\textsc{Sport}}\\ 

Metric  & AUC     & R@10     & N@10    & AUC    & R@10   & N@10  & AUC    & R@10   & N@10  & AUC    & R@10   & N@10& AUC    & R@10   & N@10\\
\midrule
BPR~\cite{bpr}    &       0.739    &      0.416          &      0.247          &       0.787        &  0.470                &     0.268    &0.897&0.699&0.442  &0.940&0.833&0.580&0.657&0.316&0.198\\
NCF~\cite{ncf}   &0.742&0.425&0.250&0.793&0.479&0.275&0.896&0.701&0.445&0.939&0.836&0.578&0.672&0.335&0.206\\
NGCF~\cite{ngcf}   &0.761&0.462&0.269&0.812&0.513&0.298&0.901&0.706&0.451&0.940&0.835&0.583&0.714&0.386&0.245\\
LightGCN~\cite{lightgcn}           &      0.808            &     0.526            &  0.324                &   0.824            &     0.560         &  0.330       &\textbf{0.903}&\textbf{0.712}&\textbf{0.456}&\textbf{0.941}&\textbf{0.841}&\textbf{0.591}&0.765&0.467&0.303\\
MostPop&-&0.358&0.208&-&0.386&0.215&-&0.527&0.312&-&0.737&0.481&-&0.366&0.216\\
\midrule
Act &0.686&0.383&0.219&0.727&0.392&0.214&0.831&0.523&0.306&0.904&0.736&0.478&0.688&0.378&0.220\\
Co-S &0.768&0.455&0.253&0.805&0.497&0.271&0.838&0.570&0.327 &0.904&0.748&0.456&0.786&0.486&0.278\\
Co-D &0.781&0.463&0.259&0.808&0.523& 0.296&0.853&0.619&0.367 &0.910&0.766&0.493&0.793&0.480&0.273\\
Int-S & 0.705 &0.356& 0.200&0.752&0.412&0.226&0.833&0.545&0.299 &0.906&0.747&0.452&0.686&0.342&0.199\\
Int-D &0.696&0.329&0.186&0.756 &0.412&0.226&0.837&0.542&0.300&0.909&0.753&0.459&0.681&0.346&0.201\\
\midrule
Int &  0.709 &0.355& 0.200 &0.755&0.413&0.227&0.844&0.552&0.321 &0.909&0.753&0.460&0.691&0.351&0.206\\
Co &    0.774& 0.462  &0.260&0.820&0.541&0.307&0.863&0.620&0.371 &0.912&0.763&0.486&0.791&0.486&0.278 \\
Act+Int  &0.722&0.406&0.231&0.760 &0.431&0.237&0.847&0.556&0.327 &0.913&0.754&0.489&0.707&0.395&0.231\\
Act+Co   &0.784&0.485&0.283&0.825 &0.549&0.315&0.865&0.621&0.374 &0.916&0.757&0.493&0.796&0.512&0.311\\
Int+Co &0.787& 0.479 & 0.277 & 0.827 & 0.549  & 0.316   &0.870&0.624&0.379&0.919&0.772&0.495&0.798&0.506&0.302 \\
Act+Int + Co  &0.790 & 0.493&0.289 &0.828&0.546&0.316 &0.871&0.622&0.378 &0.918&0.761&0.497&0.798&0.513&0.309\\
Performance $\Delta$ &$-$2.2\% &$-$7.7\%&$-$12.7\%&$+$0.5\%&$-$1.9\%&$-$4.2\%&$-$3.7\%&$-$12.4\%&$-$16.9\%&$-$2.3\%&$-$8.2\%&$-$16.2\%&$+$4.3\%&$+$9.6\%&$+$2.6\%\\
\midrule
LightGCN + F & \textbf{0.816}&\textbf{0.533}&\textbf{0.325}&\textbf{0.837}&\textbf{0.568}&\textbf{0.333}&0.902&0.703&0.450 &0.935&0.818&0.570&\textbf{0.809}&\textbf{0.535}&\textbf{0.338}  \\            \bottomrule
\end{tabular}
\caption{Recommendation results of the pre-trained recommender model on seen users and seen items, corresponding to the in-domain recommendation problem. Act denotes the activity-based feature. Co-S and Co-D represent the size-based and density-based co-occurrence features, respectively, similar to Int-S and Int-D. The Int and Co are the results after interpolation, following equation \ref{eqn:feature_mean}. The LightGCN + F is computed based on equation \ref{eqn:interpolation}, where F means the aggregated recommendation results of all the universal features (Act + Int + Co) by equation \ref{eqn:feature_interpolate}. $\Delta$ measures the performance difference between the best-performing \textit{statistical} feature combination and the best NCF model. On the Sport dataset, our proposed features outperform LightGCN by a significant margin. The reason is that  Sport is a sparse dataset. In addition, we see a smaller performance gap on sparser datasets, \eg Epinions and Yelp, compared to denser datasets, \eg MovieLens and Douban.}
\label{tab:ss_results}
\end{table*}
\renewcommand*{\factor}{0.036}

We evaluate the effectiveness of our proposed features under three recommendation problems introduced in~\cref{sec:problem_formulation} : 1) in-domain recommendation; 2) zero-shot in-domain recommendation; 3) zero-shot cross-domain recommendation.

We adopt a similar evaluation method to \cite{bpr, ncf, li2020time}. For each test interaction (user-item pair), we sample 99 unobserved items according to the user's interaction history. Then, we rank the test item for the user from the interaction among the 100 total items. We use AUC, NDCG@10, and Recall@10 as evaluation metrics on the produced rank lists. We compute the AUC score of the ranked list by giving unobserved interactions negative labels and actual test interactions positive labels. A higher AUC score indicates better ranking quality. Recall@10 measures if the given test interaction is within the top 10 out of the 100 items, and NDCG accounts for the ranking position of the test interaction. Random guessing will have an AUC score of 0.5 and a Recall@10 of 0.1. We calculate all three metrics and average the score of every test user. We repeat all the experiments five times and report the average number among the five runs.

All experiments were conducted on a Tesla V100 using PyTorch. We pre-train the recommender model for a maximum of 25 epochs using Adam optimizer. We use three feed-forward layers as the pre-trained recommendation model, with Tanh as the non-linear activation function. Same as what is being used in \cite{lightgcn, ngcf}, we fixed the embedding size to 64 for all the baselines. We train the LightGCN using two GCN layers and tune baselines in the hyper-parameter ranges centered at the author-provided values. We set the maximum output size of each feed-forward layer to 64 for a fair comparison. In addition, we tune the feature dimension $k$ in the range of $\{5, 10, 20, 50 \}$ and set the number of clusters to $k$ for both the co-occurrence-based features and interaction-based features. We use a feature dimension of 5 for all the interaction-based features on Yelp, MovieLens, and Amazon-Sport. We set the feature dimension of all the co-occurrence-based features to 20 on Yelp, MovieLens, and Douban. We choose cosine distance as the bounded distance metric $d(\cdot, \cdot)$ and euclidean distance as the similarity metric $s(\cdot, \cdot)$. All the other features' dimensions on different datasets are set to 10. 
\begin{table*}[t]
 \small
\centering
\begin{tabular}{p{0.07\linewidth}p{\factor\linewidth}p{\factor\linewidth}p{\factor\linewidth}p{\factor\linewidth}p{\factor\linewidth}
p{\factor\linewidth}p{\factor\linewidth}p{\factor\linewidth}p{\factor\linewidth}p{\factor\linewidth}p{\factor\linewidth}
p{\factor\linewidth}p{\factor\linewidth}p{\factor\linewidth}p{\factor\linewidth}}
\toprule
\textsc{Dataset} & \multicolumn{3}{l}{\textsc{Epinions}} & \multicolumn{3}{l}{\textsc{Yelp}} & \multicolumn{3}{l}{\textsc{MovieLens}}& \multicolumn{3}{l}{\textsc{Douban}}& \multicolumn{3}{l}{\textsc{Sport}}\\ 

Metric  & AUC     & R@10     & N@10    & AUC    & R@10   & N@10  & AUC    & R@10   & N@10  & AUC    & R@10   & N@10& AUC    & R@10   & N@10\\
\midrule
MostPop&-&0.371&0.217&-&0.420&0.238&-&0.510&0.294&-&0.738&0.481&-&0.370&0.213\\
\midrule
Act       &     0.696     &      0.385          &     0.223           &    0.744         &       0.425              &     0.233   &0.827&0.505&0.283 &0.904&0.724&0.474&0.686&0.383&0.218\\
Co-S    &  0.604        & 0.252          &   0.141       &      0.670          &    0.295              &  0.147     &0.768&0.453&0.248 &0.877&0.701&0.389&0.699&0.348&0.192\\
Co-D &   0.612  &  0.259  &        0.142          &  0.752     &      0.425         &   0.227 &0.777&0.437&0.235 &0.883&0.711&0.450&0.681&0.299&0.151 \\
Int-S          &   0.720            &   0.384                & 0.216                &   0.753              &     0.432       &   0.238          &0.827&0.515&0.278 &0.898&0.722&0.419&0.709&0.378&0.215\\
Int-D      &     0.690 &   0.322            &    0.179            & 0.758        &     0.432         &    0.236     &0.828 &0.511&0.272&0.887&0.662&0.350&0.693&0.372&0.214\\
\midrule
Int                         &  0.723        &  0.386              & 0.219                &             0.757     &      0.433            &     0.240           &0.833&0.519&0.296 &0.901&0.731&0.427&0.714&0.388&0.225\\
Co                        &  0.611           &   0.273                &    0.157               &        0.757          &        0.436          &     0.241        &0.799&0.483&0.269 &0.884&0.714&0.413&0.724&0.370&0.209\\
Act+Int            &0.736     & \textbf{0.422}       & \textbf{0.242}       & 0.767        &   0.450        &    0.251&0.838&0.530&0.304&0.908&0.744&0.474&0.718&0.407&0.236\\
Act+Co           &       0.695           &     0.384             &        0.222           &   0.776       &     0.464          &     0.262      &0.832&0.514&0.289 &0.902&0.731&0.465&0.733&0.423&0.247\\
Int+Co    &0.724& 0.390      &     0.221               &      0.783    &       0.478               &     0.270        &0.837&0.527&0.303&0.905&0.744&0.445&\textbf{0.753}&0.439&0.254\\
Act+I+Co       &  \textbf{0.737}&0.420         & 0.241 & \textbf{0.785}            &  \textbf{0.479}        & \textbf{0.272}      &\textbf{0.840}&\textbf{0.533}&\textbf{0.306 }&\textbf{0.909}&\textbf{0.745}&\textbf{0.478}&0.751&\textbf{0.442}&\textbf{0.258}\\
\bottomrule
\end{tabular}
\caption{\textit{Within-domain, zero-shot:} Recommendation results of the universal features on unseen users and unseen items, corresponding to the zero-shot in-domain recommendation problem. }
\label{tab:uu_results}
\end{table*}
\vspace{-5pt}
\subsection{\textbf{Experimental Results}}
\label{sec:main_results}
\subsubsection{In-domain recommendation result (\textbf{RQ1})}
In this section, we investigate the performance of our proposed feature against state-of-the-art methods under the in-domain recommendation problem (Table \ref{tab:ss_results}). Note that this experiment aims to set a baseline of state-of-the-art and universal features' recommendation performance.

Across five datasets, the individual universal features generally fall short against the selected baselines, except on the Amazon-sport dataset, where we see a 3\% improvement on AUC and 4\% on R@10 over the best-performing baseline. Among the other four datasets, with a single feature, we see a performance gap of 2\% to 5\% using AUC and 2\% to 13\% using R@10 between the best-performing universal feature and selected baseline. However, when we interpolate and combine all the features with equation \ref{eqn:feature_interpolate}, the performance gap shrinks to 3\% on AUC and 12\% on R@10. Additionally, we conduct experiments on using the features as a performance booster to arbitrary state-of-the-art recommender models via equation \ref{eqn:interpolation}. After the interpolation, we show moderate performance improvements on three of the five datasets over the best-performing baseline. In addition, we notice a trend that when the dataset is sparser, the performance difference between the proposed features and the baselines is generally smaller. We emphasize that \textit{we aim not to outperform state-of-the-art NCF models within a domain, but to lay the ground for effective zero-shot transfer by using just the interaction matrix, by developing a pre-trained recommender model}. We discuss zero-shot cases next.

\subsubsection{Zero-shot in-domain recommendation result (\textbf{RQ2})}
We empirically show the performance of the universal features on zero-shot in-domain recommendation performance (Table \ref{tab:uu_results}). We follow the procedure depicted in figure \ref{fig:model} and make predictions based on equation \ref{eqn:unseen_pred}. We train the universal model on the seen user-seen item subset of the same dataset, whose performance is in table \ref{tab:ss_results}. In general, co-occurrence-based methods have the most performance reduction from the in-domain recommendation result, ranging from 6\% to 21\% on AUC and 6\% to 40\% on recall@10. In comparison, the activity-based feature shows remarkable generalization performance, with a performance decrease on only two of the five datasets for about 0.8\% on AUC and 3\% on recall. Similar to the interaction-based feature, with maximal performance reduction of 2\% on AUC and 10\% on recall on two out of the five datasets. Note that we compute all the zero-shot in-domain recommendation results with no training for the recommendation task on the evaluated dataset. We only construct a set of universal features using part of the interactions, following the experimental setup in ~\cref{sec:setup}. We argue that these results show great potential for a truly universal recommendation system without relying on any auxiliary information.

The popularity-based baseline shows strong performance across the board. We suspect the reason is that the popularity bias is prevalent in the original datasets. Essentially, users are much more likely to interact with popular items regardless of their general preference. This can due to the popularity bias in the original recommender systems, which is already investigated in many works \cite{protocf, zhang2021causal}. However, our model can still outperform the popularity-based baseline by up to 14\%. 

\begin{table*}[t]
\centering
\begin{tabular}{lllllllllllll}
\toprule
\textsc{Dataset} & \multicolumn{3}{c}{\textsc{Douban}~\textrightarrow~\textsc{Epinions}} & \multicolumn{3}{c}{\textsc{Douban}~\textrightarrow~\textsc{Yelp}}& \multicolumn{3}{c}{\textsc{Douban}~\textrightarrow~\textsc{MovieLens}} &\multicolumn{3}{c}{\textsc{Douban}~\textrightarrow~\textsc{Sport}}\\ 

Metric  & AUC     & R@10     & N@10    & AUC    & R@10   & N@10 & AUC    & R@10   & N@10& AUC    & R@10   & N@10  \\
\midrule
Activity &0.671&0.378&0.219&0.744&0.424&0.239&0.827&0.508&0.291&0.669&0.379&0.216  \\
Co &0.604&0.261&0.150&0.677&0.359&0.199&0.798&0.463&0.255&0.626&0.248&0.129 \\
Int&0.691&0.390&0.226&0.753&0.444&0.254&0.830&0.514&0.294&0.711&0.398&0.232 \\
Act+Co&0.670&0.377&0.218&0.744&0.424&0.239&0.827&0.507&0.287&0.681&0.381&0.218 \\
Act+Int&0.695&\textbf{0.397}&\textbf{0.229}&0.755&0.438&0.248&0.835&0.522&0.300&0.704&0.394&0.226 \\
Co+Int&0.693&0.388&0.224&0.751&0.432&0.245&0.834&0.522&0.299&0.719&0.403&0.232 \\
Act+I+Co&\textbf{0.696}&0.394&0.227&\textbf{0.757}&\textbf{0.441}&\textbf{0.249}&\textbf{0.834}&\textbf{0.522}&\textbf{0.301}&\textbf{0.709}&\textbf{0.395}&\textbf{0.227} \\
\midrule
\midrule
 \textsc{Dataset}&\multicolumn{3}{c}{\textsc{Epinions}~\textrightarrow~\textsc{Douban}} & \multicolumn{3}{c}{\textsc{Epinions}~\textrightarrow~\textsc{Yelp}}& \multicolumn{3}{c}{\textsc{Epinions}~\textrightarrow~\textsc{MovieLens}} &\multicolumn{3}{c}{Epinions~\textrightarrow~Sport}\\ 
Metric  & AUC     & R@10     & N@10    & AUC    & R@10   & N@10 & AUC    & R@10   & N@10& AUC    & R@10   & N@10  \\
 \midrule
Activity &0.861&0.552&0.335&0.739&0.424&0.236&0.767&0.346&0.186&0.691&0.383&0.217 \\
Co &0.848&0.705&0.376&0.721&0.405&0.224&0.767&0.459&0.244&0.618&0.256&0.133 \\
Int&0.815&0.456&0.246&0.760&0.416&0.224&0.759&0.346&0.174&0.717&0.394&0.228 \\
Act+Co&0.885&0.708&0.412&0.751&0.436&0.245&0.780&0.455&0.246&0.696&0.385&0.219 \\
Act+Int&0.859&0.643&0.387&0.766&0.454&0.250&0.809&0.444&0.245&0.721&0.405&0.234 \\
Co+Int&0.861&0.720&0.414&0.772&0.452&0.250&0.798&\textbf{0.499}&\textbf{0.274}&0.721&0.404&0.230 \\
Act+I+Co&\textbf{0.879}&\textbf{0.724}&\textbf{0.442}&\textbf{0.773}&\textbf{0.461}&\textbf{0.247}&\textbf{0.814}&0.493&0.271&\textbf{0.724}&\textbf{0.409} &\textbf{0.235}\\
\bottomrule
\end{tabular}
\caption{\textit{Cross-domain, zero-shot:}  Recommendation results of cross dataset transfer on unseen users and unseen items. $A \rightarrow B$ means we pre-train the recommender system using A's data and directly test on B. We notice that recommender model pre-trained on denser dataset generalize better on denser dataset (\textsc{Douban}~\textrightarrow~\textsc{MovieLens} vs. \textsc{Epinions}~\textrightarrow~\textsc{MovieLens}. Similarly, recommender model pre-trained on sparse dataset generalize better to sparse datasets.}
\label{tab:cu_results}
\end{table*}

  \vspace{-5pt}
\subsubsection{Zero-shot cross-domain recommendation result (\textbf{RQ3})}
We present the result of the zero-shot cross-domain recommendation problem introduced in ~\cref{sec:problem_formulation}. Like the zero-shot in-domain recommendation, we do not train for the recommendation task but use part of the available interaction matrix to construct universal features. The difference is that we use the data from another domain to pretrain the recommender model. We use the same test set as the zero-shot in-domain recommendation problem for evaluating the result. The activity-based feature again shows excellent generalization performance across the board, with exceptions when transferring the pretrained transformation layer from Epinions to Douban and MovieLens. We argue that the most probable reason is that these datasets are much denser than others. Therefore, their activity distribution may exhibit different characteristics. Interaction-based features also show outstanding generalization performance across the dataset, while co-occurrence-based features have the worst generalizability. When interpolating the results, we observe a maximum reduction of 9\% on AUC and 10\% on recall. These experimental results empirically show that the proposed features are universal across datasets, and the pretrained universal model is highly transferrable.

\subsection{\textbf{Qualitative Analysis (RQ4)}}
\label{sec:analysis}
In this section, we try to answer how will the amount of training data influences the zero-shot generalization performance. We fix the test set and use the zero-shot in-domain recommendation as our testing scenario. First, we vary the percentage of interaction used as training for each user. This analysis examines the effect of the amount of training data points, while maximally preserving the original data distribution by sampling a fixed percentage of data per user. Next, we will look at the effect of the percentage of users and items (seen percentage) used to construct the in-domain training set. Lowering the seen percentage results in a nosier dataset in terms of data distribution.

   \begin{figure*}[t]
     \centering
      \includegraphics[width=\linewidth]{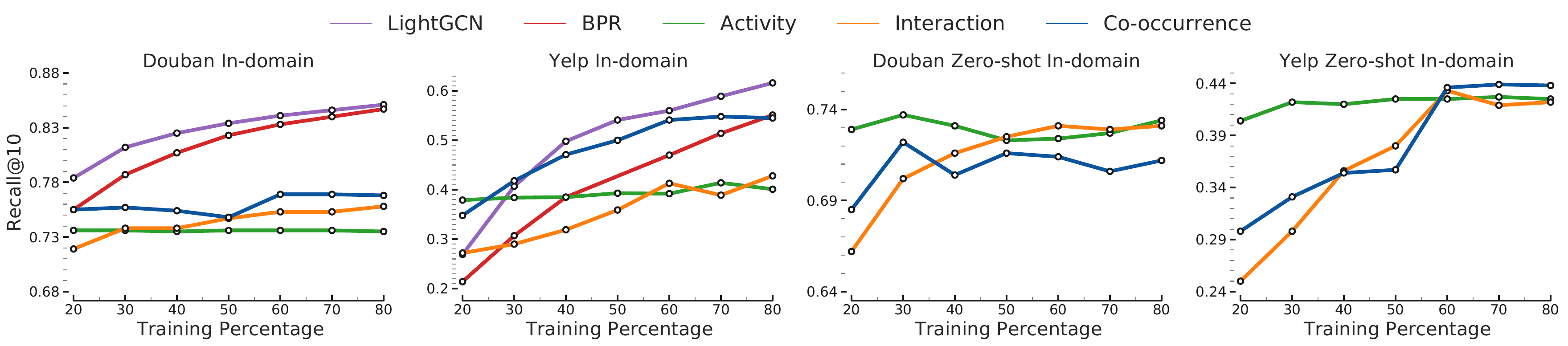}
      \caption{In-domain and zero-shot in-domain recommendation result vs training percentage of each user's interaction history. We fix both the in-domain and the zero-shot in-domain test set. For each user, we vary the percentage of interactions for feature construction and training. When the training percentage is 60, it corresponds to experimental results in Table ~\ref{tab:ss_results} and Table ~\ref{tab:uu_results}.}
      \label{fig:vary_cold}
  \end{figure*}
\vspace{-8pt}

\subsubsection{Results on varying training interaction percentage}
From Figure \ref{fig:vary_cold}, we can observe that both in-domain and zero-shot in-domain performance on the Douban dataset converges quickly with the increase in training interaction percentage per user, whereas Yelp takes much longer. In general, we see a gradual increase in both in-domain and zero-shot in-domain performance when we increase the training percentage per user. With lower training percentages, \ie 20\% and 30\%, we see that the gap in in-domain recommendation performance between state-of-the-art models and the proposed features rapidly decreased. In addition, the activity-based feature is less likely to be influenced by the training percentage, whereas the other features show changes in results. 

    \begin{SCfigure}[][h]
    {
    \includegraphics[width= 80mm]{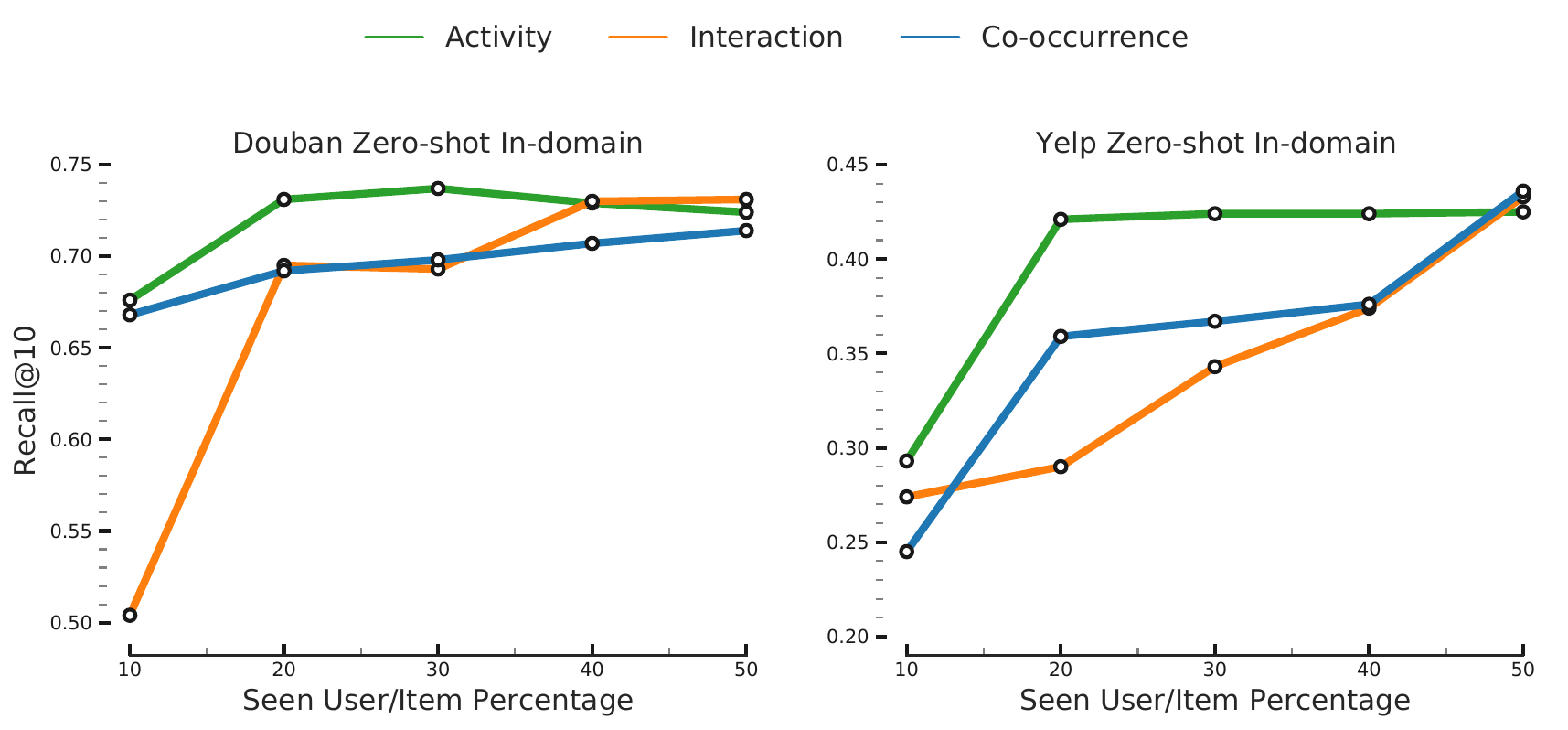}
        \captionsetup{width=\linewidth}
        \caption{Zero-shot in-domain recommendation result v.s the seen percentage of users and items. We fix the zero-shot in-domain test set and keep the seen percentage of users and items the same. The x-axis denotes how many percentages of the users and items are used to construct the seen user-seen item dataset used for training. When the seen user/item percentage is 50, it corresponds to the results presented in Table ~\ref{tab:ss_results} and Table ~\ref{tab:uu_results}. We can see that the performance of all the features converges around 20 seen percentage on Douban. The activity-based feature converges the quickest among all the datasets.}\label{fig:vary_train}
        }
    \end{SCfigure}
\subsubsection{Results on varying seen percentage}
As shown in Figure \ref{fig:vary_train} we observe a considerable performance increase when increasing the seen percentage from 10 to other numbers. It is because when only using 10 \% of users and items of the original dataset, the data distribution of the training dataset will be very different from the zero-shot in-domain test set. In addition, the plot on the Douban dataset converges much quicker than that on Yelp, which probably contributes to the fact that Douban is a much denser dataset. Therefore, we can recover the original distribution with even a small percentage of the users and items. Whereas on Yelp, the performance gradually converges until we use 40\% of the users and 40\% of the items as the seen percentage.

\subsection{\textbf{Parameter Sensitivity (RQ5)}}
\label{sec:model_analysis}
   \begin{figure*}[t]
     \centering
      \includegraphics[width=\linewidth]{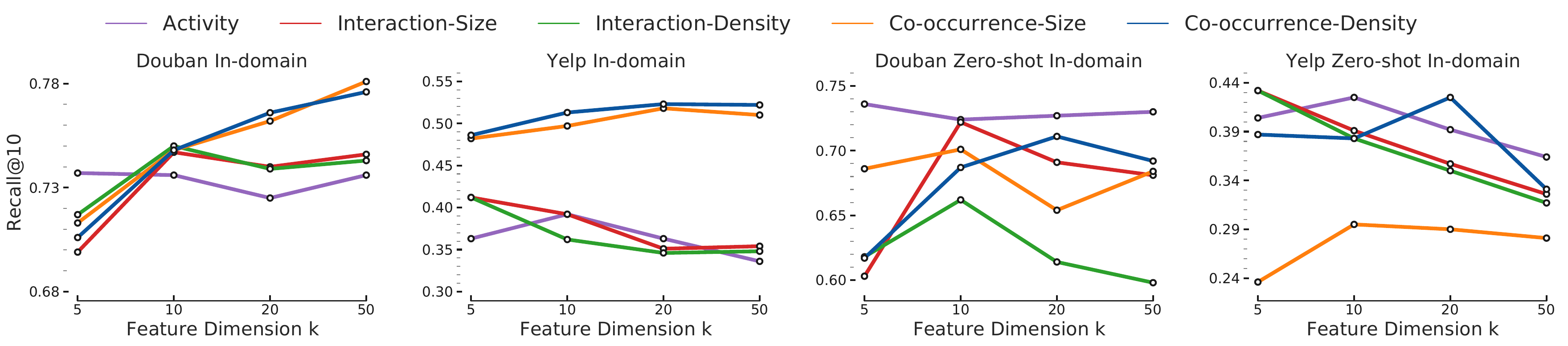}
      \caption{In-domain and zero-shot in-domain recommendation result vs feature size $k$. The x-axis denotes the feature size/bin size $k$ and the y-axis represents the performance, measured by Recall@10. All dataset-agnostic features show performance decrease on the zero-shot in-domain recommendation task as $k$ increases. The co-occurrence-based features benefit from a larger $k$ size for the in-domain recommendation problem. These features achieve optimal performance on both recommendation problems in different $k$ values.}
      \label{fig:sensitivity}
  \end{figure*}
We analyze sensitivity to the number of bins (feature dimension) $k$ in the feature construction. In Figure \ref{fig:sensitivity}, we show the effect of different $k$ on in-domain and zero-shot in-domain recommendation performance. While larger $k$ for both co-occurrence features correspond to better in-domain recommendation performance, the zero-shot in-domain performance decreases. When $k = 10$, the size-based co-occurrence feature has the best overall performance across the selected datasets, while the density-based co-occurrence feature achieves optimal performance when $k = 20$. On the other hand, both the size and density-based interaction features achieve the most excellent performance when $k = 10$ and $k = 20$ on Douban and Yelp, respectively. Finally, the activity-based feature generally achieves the best overall performance on the in-domain and zero-shot in-domain recommendation when $k = 10$.
\vspace{-5pt}
\subsection{\textbf{Discussion}}
\label{sec:discussion}
Unlike previous works that use auxiliary information, our pre-trained recommender model shows excellent potential for zero-shot recommendation with only interaction data. We construct dataset-agnostic universal features based on the behavior of the users and items and the statistical characteristics of the interaction data, on top of which we develop a simple yet effective pre-trained recommendation model that can generalize to unseen users and items within a dataset and across different datasets/domains.

While the proposed framework shows great potential as a pre-trained recommender model, it has several weaknesses. Firstly, the parameters related to the feature construction process are selected arbitrarily. In contrast, we could model the distributions by their cumulative distribution function (CDFs) in future works. In addition, adopt Node2Vec to capture the co-occurrence patterns of users and items, which requires additional preprocessing time and has to be performed on every dataset. We could potentially represent each node not by its ID but by where they lie in the CDFs. However, with the extensive experiments conducted, this work pioneers the pre-trained recommender model, which can inspire future works in this direction. 
\vspace{-10pt}

\section{Conclusion}
\label{sec:conclusion}

Inspired by the impact of pre-trained models, we explored the possibility of pre-trained recommender models that support building recommender systems in new domains, with minimal or no retraining, without the use of any auxiliary user or item information. Our insight was that the statistical characteristics of the user-item interaction matrix will be critical in learning universal representations for users and items. We showed how to learn universal (\ie supporting zero-shot adaptation without user or item auxiliary information) representations for nodes and edges from the bipartite user-item interaction graph. We learned the representations by exploiting the statistical properties of the interaction data, including user and item marginals, and the size and density distributions of their clusters. We show that a simple pre-trained model consisting of several feed-forward neural layers, combined with proposed features, can achieve excellent generalization performance on unseen users and items (\ie zero-shot setting) within and across different datasets/domains. In addition, we show that the proposed features can also boost the performance of state-of-the-art neural recommenders by up to 14\% in the traditional recommendation setting.



\bibliographystyle{ACM-Reference-Format}
\bibliography{main,hs}

\end{document}